\documentstyle[12pt]{article}

\makeatletter
\@addtoreset{equation}{section}
\makeatother

\def\ba{\begin{eqnarray*}}
\def\ea{\end{eqnarray*}}
\def\ban{\begin{eqnarray}}
\def\ean{\end{eqnarray}}

\def\@normalsize{\@setsize\normalsize{10pt}\xpt\@xpt
\abovedisplayskip 10pt plus2pt minus5pt\belowdisplayskip \abovedisplayskip
\abovedisplayshortskip \z@ plus3pt\belowdisplayshortskip 6pt plus3pt
minus3pt\let\@listi\@listI}

\def\subsize{\@setsze\subsize{9pt}\xipt\@xipt}

\def\subsection{\@startsection
{subsection}{2}{\z@}{9pt plus 0.3pt minus 0.3pt}
{9pt plus 0.3pt minus 0.3pt}{\small\bf}}

\makeatother
\newcommand{\D}{\displaystyle}

\newcommand{\bq}[1]{\begin{equation}\label{#1}}
\newcommand{\eq}{\end{equation}}
\newcommand{\refq}[1]{(\ref{#1})}

\newcommand{\intli}{\:\!\!\!\!\int\limits }

\newcommand{\defeq}{\:\stackrel{\rm def}{=}\:}

\newcommand{\expj}[1]{e^{-j2{\pi}{#1}}}
\newcommand{\expjp}[1]{e^{j2{\pi}{#1}}}

\def\ni{\noindent}

\begin{document}
\bibliographystyle{plain}
\date{}

\title{\large\bf QUADRATIC TIME--VARYING SPECTRAL ESTIMATION\\[2mm] 
FOR UNDERSPREAD PROCESSES\thanks{Funding by grant 4913 of the 
                      {\em Jubil{\"a}umsfonds der {\"o}sterreichischen
                      Nationalbank and the US Dept.~of Energy.}\/}
}
\author{
 \normalsize Werner Kozek \\[1.5mm]
 \normalsize NUHAG, Dept.~of Mathematics\\ 
 \normalsize  University of Vienna\\
 \normalsize  Strudlhofg.~4, A--1090 Vienna, Austria\\
 \normalsize  (kozek@tyche.mat.univie.ac.at)
\and 
\normalsize Kurt Riedel\\[1.5mm]
 \normalsize Courant Institute of Mathematical Sciences\\
 \normalsize   New York University\\
 \normalsize New York, New York 10012\\
 \normalsize (riedel@cims.nyu.edu)
}
\date{\vspace*{-3ex}}     
\maketitle
\thispagestyle{empty}

\small

\noindent{\bf Abstract:} Time--varying spectral estimation is studied
for nonstationary processes with restricted time--frequency (TF)
correlation.
Explicit bias and variance expressions are given for quadratic 
TF--invariant (QTFI) estimators of an
expected real--valued QTFI representation based on a single noisy
observation. Unbiased theoretical estimators
with globally minimal variance are derived and 
approximately realized by a matched multi-window method.

\vspace{0.5mm}

\section{INTRODUCTION}  \vspace*{-1mm}

Time-frequency distribution are widely used to search for hidden structure
in the signal. When the signal consists of a small number of slowly varying
sinusoids, the Wentzel-Kramer-Brillioun representation reduces the signal
to curves in the time-frequency plane \cite{rie94}. 
We consider the case of nonstationary
stochastic processes with underlying time-frequency structure in the
correlation operator.

The evolutionary spectrum is one common representation of nonstationary
processes. In \cite{rie93}, 
we propose estimating the evolutionary spectrum by
smoothing the log-spectrogram using a data-adaptive kernel smoother in
the time-frequency plane. The evolutionary spectrum has two advantages: it
is always positive and it converges to the spectrum as the ratio of the
characteristic time scale to the sampling rate becomes large. Its
disadvantages are its lack of uniqueness
and its relatively poor time frequency resolutions.

We consider a different class of representations of nonstationary processes:
quadratic Cohen's class spectra. These representations correspond to
the expected value of Cohen's class time-frequency representations. An
important member of this class is the Wigner-Ville spectrum. This class of
spectral representations possess useful operator properties and a
reproducing kernel Hilbert space structure.

In this article, we consider the estimation problem: how to estimate the
Cohen's class spectra. This same problem has been considered by Sayeed
and Jones as well. In \cite{say94}, a complete knowledge of the correlation
operator is assumed. This assumption is appropriate for the signal
classificiation problem of recognizing one or more specific signals. Our
approach assumes much weaker {\it a priori} knowledge. We assume only that
the signal is underspread which corresponds to being double band limited
in the ambiguity plane.

In Section 2, we review time-frequency representations of deterministic
signals. Section 3 presents the analogous theory for time varying spectra.
Section 4 defines and motivates underspread processes.

Section 5 analyzes the bias and variance of a special class of quadratic
estimators of Cohen's class spectra. Section 6 determines the minimum
variance unbiased estimator of an underspread process. Section 7 describes
a related estimation using multiple windows. Section 8 presents a
biomedical example.

\section{QUADRATIC TIME-FREQUENCY DISTRIBUTIONS}

\noindent Every real--valued quadratic time--frequency (TF) shift--invariant
(QTFI) representation of a signal $x(t)$ can be represented
as a quadratic form \cite{coh66}.
Comprehensive reviews of Cohen's class are given in \cite{coh89,Hla92r}.
We now cast Cohen's class of time-frequency representations in an operator
theoretic framework. Let  ${\bf P}$ be a self--adjoint Hilbert-Schmidt (H-S)
``prototype'' operator. We define the 
quadratic time-frequency shift invariant (QTFI) distribution
\bq{detspec}
T_x(t,f)=\big< {\bf P}^{(t,f)}x,x \big> ,
\eq
where ${\bf P}^{(t,f)}$ is a TF shifted version 
of the self--adjoint prototype operator ${\bf P}$.
The choice of
the kernel, $p(s,t)$, determines a particular representation in Cohen's class.
The TF--shifting of operators is defined as
$
   {\bf P}^{(t,f)}\defeq{\bf S}^{(t,f)}{\bf P}{\bf S}^{(t,f)+},
$
where ${\bf S}^{(\tau,\nu)}$ is a unitary TF shift operator, 
acting as
$
\left({\bf S}^{(\tau,\nu)}x\right)(t)=x(t-\tau)\expj{\nu t}.
$
The  standard H-S inner product, $< {\bf R,P}>$, is 
$$
<{\bf R,P}>=\int\int r(t,s)p(s,t)dtds \ ,
$$
where $ r(t,s)$ and $p(s,t)$ are the respective kernels of the 
H.-S. operators  ${\bf R}$ and ${\bf P}$.
Throughout this article, we assume an infinite time domain and suppress
replace $\int_{-\infty}^{\infty} dt$ with $\int dt$.

We now review
the basic unitary TF representations of HS operators \cite{koz92a}.
The generalized Weyl symbol is defined as
$$ L^{(\alpha)}_H(t,f)
\defeq \!\int_t h
\!\left(\!t\!+\!\left(\!\frac{1}{2}\!-\!\alpha\!\right) \!
         \tau, t\!-\!\left(\!\frac{1}{2}\!+\!\alpha\!\right) \!\tau\!\right) 
       \expj{f \tau } d\tau \vspace{-1mm},
$$
where $|\alpha |\le 1/2$. The Weyl correspondence is given by $\alpha=0$,
and the Kohn-Nirenberg correspondence 
(time--varying transfer function) by $\alpha=1/2$
\cite{she94}. (When
we suppress the superscript, this means validity for any $\alpha$.)
The TF shifting of operators corresponds to a shift of the symbol,
$
L_{P^{(\tau,\nu)}}(t,f)=L_P(t-\tau,f-\nu),
$
which shows that whenever ${\bf P}$ is a
TF localization operator that selects signals centered in the
origin of the TF plane then ${\bf P}^{(t,f)}$  localizes
signal components centered around $(t,f)$.
The {\em generalized spreading function} (GSF)
of a linear operator \cite{koz92a} is
$$
S^{(\alpha)}_H(\tau,\nu)\! \defeq \!\!\int_t \!h
\!\left(\!t\!+\!\left(\!\frac{1}{2}\!-\!\alpha\!\right) \!
         \tau, t\!-\!\left(\!\frac{1}{2}\!+\!\alpha\!\right) \!\tau\!\right) 
     \!  \expj{\nu t}\! dt.
$$
The GSF is the symplectic Fourier transform of the generalized
Weyl symbol $L_H^{(\alpha)}(t,f)$:
\bq{twod}
S^{(\alpha)}_H(\tau,\nu)
      =\intli_t\intli_f L_H^{(\alpha)}(t,f)
      \expj{(\nu t-\tau f)}dtdf \ ,
\eq
\bq{twodb}
 L_H^{(\alpha)}(t,f)
      =\intli_{\tau}\intli_{\nu}
S^{(\alpha)}_H(\tau,\nu)      \expj{(-\nu t+\tau f)}d\tau d\nu \ .
\eq
When the Weyl symbol is smoothly varying in time
and frequency, then the generalizing
spreading function decays in $\tau$ and $\nu$.

\section{TIME VARYING SPECTRUM}

\ni For a nonstationary process, a 
time--varying spectrum may be defined as the expectation
of \refq{detspec}\vspace*{-0.5mm}  
\bq{stospec}
P_x(t,f)\defeq\mbox{E}\left\{\big<{\bf P}^{(t,f)}x,x\big>\right\},
\eq
A prominent example for $P_x(t,f)$ is the Wigner--Ville spectrum
\cite{mar85}.  Priestley's evolutionary
spectrum \cite{pri65, rie93} is a different, popular definition of
a stochastic time--varying spectrum that 
cannot be brought into the form of \refq{stospec}.

We consider {\em circular} complex, zero--mean 
Gaussian processes with trace--class correlation kernel
$$
\left({\bf R}_x\right)(t,t')=r_x(t,t')=\mbox{E}\left\{x(t)x^*(t')\right\},
\hspace*{5mm}
\mbox{tr}{\bf R}_x<\infty.
$$
The trace--class convention implies a
HS inner product representation of $P_x(t,f)$, 
alternatively written as the trace of the product operator:
\bq{spedef2}
P_x(t,f)=\big < {\bf R}_x,{\bf P}^{(t,f)}\big>
=\mbox{tr}\left\{{\bf R}_x{\bf P}^{(t,f)}\right\}.
\eq

The {\em expected ambiguity function} is defined as the GSF of the correlation
operator \cite{koz94b}\vspace*{-0.5mm}
\bq{eadef} 
EA^{(\alpha)}_x(\tau,\nu) \defeq  S^{(\alpha)}_{R_x}(\tau,\nu)
\eq
With the generalized Wigner--Ville spectrum, defined as\vspace*{-0.5mm}
$$ 
EW^{(\alpha)}_x(t,f) \defeq L^{(\alpha)}_{R_x}(t,f),
$$
Eq.~\refq{twod} carries over to
a ``nonstationary Wiener--Khintchine relation'':\vspace*{-1.5mm}
\bq{wiki}
 EW_x^{(\alpha)}(t,f)= 
\int_\tau\int_\nu
 EA_x^{(\alpha)}(\tau,\nu) \expjp{(\nu t-\tau f)} 
 d\tau\,d\nu\ , \vspace*{-0.5mm}
\eq
\bq{wikib}
 EA_x^{(\alpha)}(\tau,\nu) =
\int_t\int_f
 EW_x^{(\alpha)}(t,f)= 
\expjp{(\tau f-\nu t)} 
 dt\,df\ . \vspace*{-0.5mm}
\eq
These relation ships are summarized in Table 1.

\ \\

\begin{tabular}{lcl}
$L_R^{(\alpha )} (t,f)$ & = & $EW^{(\alpha )} (t,f)$ \\
Weyl symbol of correlation & & Generalized W-V spectrum \\
$ \ \ \ $ $\Uparrow$ $t \leftrightarrow \nu$&  
&$\Uparrow$ $t \leftrightarrow \nu$  \\
$ \ \ \ $ $\Downarrow$ $f \leftrightarrow \tau$&  
&$\Downarrow$ $f \leftrightarrow \tau$ \\
$S_R^{(\alpha )} (t,\nu)$ & = & $ EA(\tau ,\nu )$ \\
GSF of correlation & & Expected ambiguity function
\end{tabular}

Table 1

\ \\

As an example, the real--valued generalized
Wigner--Ville spectrum can be written as
$$
\mbox{Re}\left\{EW_x^{(\alpha)}(t,f)\right\}=
\mbox{E}\left\{\big<{\bf P}^{(t,f)}\!(\alpha)\,x,x\big>\right\} \ ,
$$
where the $\alpha$--dependent prototype operator is given by:\vspace*{-1mm}
\bq{gwvsp} 
S^{(0)}_{P(\alpha)}(\tau,\nu)=\cos(2\pi\tau\nu\alpha)  \ .
\eq

Since both the Weyl symbol and the spreading function
are unitary representations of HS operators we
can rewrite the general time--varying spectrum,\vspace*{-0.5mm} 
$$
\big < {\bf R}_x,{\bf P}^{(t,f)}\big>=
\big < EW_x,L_{P^{(t,f)}}\big>=
\big < EA_x, S_{P^{(t,f)}}\big> \ .
$$

Note furthermore that the GSF of the TF shifted
prototype operator is just a modulated version
of the GSF of the original version:\vspace*{-1.5mm}
$$
S_{P^{(t,f)}}(\tau,\nu)=S_P(\tau,\nu)\expjp{(\nu t-\tau f)},
$$
thus in particular $|S_{P^{(t,f)}}(\tau,\nu)|= |S_P(\tau,\nu)|$.

\section{UNDERSPREAD PROCESSES} 

\noindent The bias-variance analysis of Sec.~5 is valid for any 
circular Gaussian process with  a trace class covariance.
We now restict our consideration to the case where  
the process' expected ambiguity function 
$EA^{(\alpha)}_x(\tau,\nu)$ is zero outside a rectangle
in the ambiguity plane.
Our requirement that the expected ambiguity is double band-limited implies
that the Weyl symbol is smooth in time and frequency.

We denote the maximum temporal correlation width $\tau_{\max}$
and the  maximum spectral correlation width $\nu_{\max}$;
i.e., we assume that the expected ambiguity function satisfies
\bq{sprecon}
EA_x^{(\alpha)}(\tau,\nu)=EA_x^{(\alpha)}(\tau,\nu)\chi_x(\tau,\nu),
\eq
where $\chi_x(\tau,\nu)$ is the $0/1$--valued 
indicator function of a 
centered rectangle with area $s_x=4\tau_{max}\nu_{max}$. 
According to the recently
introduced terminology we call a process with
$s_x<1$ {\em underspread} and in the converse case {\em overspread} 
\cite{koz93a}. 
For asymptotics we assume that the underspread parameter is very small:
$s_x \ll 1$. The underspread parameter, $s_x$, corresponds to the
expansion parameter $1/(\tau\lambda_f )$, which is used in the analysis of
evolutionary spectra \cite{rie93}.

As to the relevance and realizability of the underspread processes
we note that practically important linear time--varying (LTV) systems,
as e.g. the mobile radio channel or underwater acoustic channel \cite{sos68},
are characterized by an (at least in good approximation) 
restricted spreading function (this is the field where
the underspread/overspread terminology was originally introduced).
Now, we apply stationary
white noise $n(t)$ with $\mbox{E}\{n(t)n^*(t')\}=\delta(t-t')$
to an underspread LTV system ${\bf H}$ characterized by\vspace*{-0.5mm}
$$
S^{(\alpha)}_H(\tau,\nu)=S^{(\alpha)}_H(\tau,\nu)\chi_H(\tau,\nu)
\vspace*{-0.5mm}
$$
where $\chi_H(\tau,\nu)$ covers a centered rectangle 
with halfwidths $\tau_{max,H}$ and $\nu_{max,H}$. 
Then the output process $x(t)=({\bf H}n)(t)$ is nonstationary
with correlation \vspace*{-1mm}
$$
{\bf R}_x ={\bf H}{\bf H}^+.
$$
Applying the triangle inequality to the spreading function of
the product operator \cite{koz94d} gives \vspace*{-1.5mm}
$$
 |EA_x(\tau,\nu)| \le |S_H(\tau,\nu)|** |S_H(\tau,\nu)|,
$$
where the $**$ denotes double convolution.
The output process is thus underspread with
$\tau_{max,x}=2\tau_{max,H}$ and $\nu_{max,x}=2\nu_{max,H}$.
Hence, we have shown that underspread processes are realizable
and relevant.

In view of the ``nonstationary Wiener--Khintchine relation''
\refq{wiki},  the
overspread/\-underspread classification may be interpreted as a smoothness 
condition for the time--varying spectrum of the process.
Applying the sampling theorem on the symbol level
leads to a {\em discrete Weyl--Heisenberg expansion}\/ of the correlation
operator \cite{koz94d}:\vspace*{-1.5mm}
$$
  {\bf R}_x=\sum_l\sum_m  EW_x^{(\alpha)}(lT,mF) 
   {\bf P}^{(lT,mF)}(\alpha) \vspace*{-0.5mm} 
$$
valid for a sampling grid with\vspace*{-1.5mm}
$$
  T\le\frac{1}{2\nu_{max}}\hspace*{1cm}\mbox{and}\hspace*{1cm}
   F\le\frac{1}{2\tau_{max}}
$$
and the prototype operator defined by
\bq{Sqr}
S^{(\alpha)}_{P(\alpha)}(\tau,\nu)=\chi_x(\tau,\nu).\vspace*{-0.5mm}
\eq
The critical spread $s_x=1$ corresponds to the Nyquist
sampling density $TF=1$. Hence, considering bandlimited
processes, for  $s_x=1$ the rate of innovation
in the process second order statistics is equal to
the sampling rate of the realization \cite{koz94b}.
However, a robust estimation procedure maps a time series
with $N$ samples on a model with less than $N$ coefficients
such that the critical spread is a treshold for robust
estimation of the generalized Wigner--Ville spectrum.
It is furthermore remarkable that the evolutionary
spectrum of an underspread process is 2D bandlimited
in exactly the same manner as the generalized 
Wigner--Ville spectrum \cite{koz94d}. 

It should be noted
that one can view the stationarity assumption underlying any
time--invariant spectrum estimation as 
a limit case of  \refq{sprecon} since the expected
ambiguity function of a wide--sense stationary
process is characterized by ideal concentration
on the $\tau$--axis:\vspace*{-1mm}
$$
EA_x(\tau,\nu)=r_x(\tau)\delta(\nu),\vspace*{-1mm}
$$
where $r_x(\tau)$ is the autocorrelation function.

The Wiener--Khintchine relation requires 
strict band-limiting the ambiguity plane.
The remainder of our analysis requires only a concentration in the
ambiguity plane with characteristic spread, $s_x \ll 1$, but not
complete band limitation.

\section{REPRODUCING KERNEL HILBERT SPACE} 

We now show that time-frequency distributions are a  
reproducing kernel Hilbert spaces (RKHS) \cite{kai73}
using the Wigner-Ville kernel.
A RKHS is Hilbert space ${\cal H}$
of complex valued functions, defined on a set ${\cal S}$, that has
a reproducing kernel $K(s,t)$  defined on  ${\cal S}\times{\cal S}$
with two properties: (i) for each t, the function K(s,t) lies in
${\cal H}$  and (ii) for each $x\in{\cal H}$ and each $t\in {\cal S}$
one has the reproducing property: 
$$
x(t)=\big< x, K(.,t)\big>=\intli_{t'} K^*(t',t)x(t') dt'.
$$
In our case, the Hilbert space,  ${\cal H}$, is
the set of
 Weyl symbols of underspread operators  ${\bf H}$
satisfying a given spreading constraint \refq{sprecon}.
The reproducing kernel is given by
the Weyl symbol of the prototype operator:
$$
K(t',f',t,f)=L_{P^{(t,f)}}(t',f').
$$
This is in fact a reproducing kernel 
as (i) for each $(t,f)$ $P^{(t,f)}$ remains underspread
since
$$
\left|S_{P^{(t,f)}}(\tau,\nu)\right|=\left|S_P(\tau,\nu)\right|,
$$
and (ii) one has the reproducing formula as follows:
\bq{RKHSeq}
 L_H(t,f)
 =\big< L_H, L_{P^{(t,f)}}\big>=
 \intli_{t'}\intli_{f'} L_H(t',f') L_P(t'-t,f'-f) dt'\,df'
\eq

WERNER: DO YOU MEAN \ref{RKHSeq} for the Wigner Ville
kernel or for the kernel in\ref{Sqr} or WHAT?  
 
\section{ QTFI ESTIMATION} 

\ni
We now consider QTFI estimators of the time varying spectrum 
of the signal process $x(t)$ when it is contaminated with noise.
We are given  a single noisy observation, $y(t)$ of the signal 
process $x(t)$: \vspace*{-0.5mm}
$$
y(t)=x(t)+n(t) \hspace*{5mm}\mbox{with}\hspace*{5mm}
\mbox{E}\left\{n(t)n^*(t')\right\}=\sigma_n^2\delta(t-t'),
$$
where $n(t)$ is statistically independent, zero--mean, {\em circular} complex 
Gaussian white noise. To estimate $P_x(t,f)$
we use a generally  different QTFI representation
of the observation: 
$$
{\widehat P}_x(t,f)=\big<{\bf \widehat P}^{(t,f)}y,y\big>.
$$
We define the ``bias operator''
 ${\bf\widetilde P}$ as 
$
{\bf\widetilde P}\defeq{\bf\widehat P}-{\bf P}.
$
The QTFI estimator is consistent with
classical, ``non--parametric'' time--invariant spectrum estimation
where the predominant class of estimators  \cite{tho82,RST94}
can be basically written
as a frequency parametrized quadratic form:
\bq{tinsp}
{\hat S}_x(f)=\big< {\bf\widehat P}^{(0,f)} y,y\big> \ . \vspace*{1mm}
\eq
  
\section{BIAS AND VARIANCE ANALYSIS} 
With the statistical
independence of signal and noise  and using \refq{spedef2}
we have the following expectation of the estimate:
\bq{mue}
\mbox{E}\left\{{\widehat P}_x(t,f)\right\}
=\mbox{E}\left\{\big<{\bf \widehat P}^{(t,f)}x,x\big>\right\}+
\sigma^2_n\mbox{tr}{\bf \widehat P} \ ,
\eq
such that the bias is given by
$$
B(t,f)\defeq \mbox{E}\left\{{\widehat P}_x(t,f)\right\}- P_x(t,f)=
\mbox{tr}\left\{{\bf \widetilde P}^{(t,f)}{\bf R}_x\right\}+
\sigma_n^2\mbox{tr}{\bf\widehat P}.
$$
Using the Schwarz inequality for operator inner products and triangle
inequality, we immediately get a tight bound for the maximum bias:
\bq{bmax}
    |B(t,f)|\le\|{\bf\widetilde P}\|\|{\bf R}_x\|+
    \sigma_n^2 |\mbox{tr}{\bf\widehat P}| \ ,
\eq
where the operator norm is the HS norm.
We assume knowledge of the noise level $\sigma_n^2$ such that we can trivially
correct the TF--independent  bias term:
$$
 {\widehat P'}_x(t,f)=
 {\widehat P}_x(t,f)-\sigma_n^2\mbox{tr}{\bf\widehat P},
$$
where ${\widehat P'}_x(t,f)$ denotes the corrected estimate.

The variance,\vspace*{-1mm} 
$$ 
V(t,f)\defeq\mbox{E}\left\{{\widehat P}^2_x(t,f)\right\}-
{\left(\mbox{E}\left\{{\widehat P}_x(t,f)\right\}\right)}^2,
$$
is evaluated using of Isserlis' fourth order
moment formula (for the special case of circular complex variables),
$
 \mbox{E}\left\{x(t_1)x^*(t_2)x(t_3)x^*(t_4)\right\}=
r_x(t_1,t_2)r_x(t_3,t_4)
 + r_x(t_1,t_4)r_x(t_3,t_2),
$ one has:
$$
V(t,\!f)\!=\!
\mbox{tr}\!\left\{\!\!{\left({\bf\widehat P}^{(t,f)}{\bf R}_x\right)}^2\!
\right\}
\!+2\sigma_n^2\mbox{tr}\!\left\{\!{\left({\bf \widehat P}^{(t,f)}\right)}^2
{\bf R}_x\!\!\right\}+\sigma_n^4\|{\bf \widehat P}\|^2\!\!.
$$
The Schwarz inequality for the operator inner product leads to
a bound on the maximum variance,
\bq{vmax}
V_{max}\le{\|{\bf\widehat P}\|}^2\left(\|{\bf R}_x\|+\sigma_n^2\right)^2, 
\eq
proportional to the HS norm of the prototype operator
${\bf\widehat P}$.

\newpage
\noindent{\bf Global Mean Square Error.}
The bias and variance results are complicated TF--dependent expressions.
Due to our restriction to QTFI estimators we need
TF--invariant, thus global indicators for the estimator performance.
After correcting for the  TF independent bias term,
$B_0\defeq \sigma_n^2\mbox{tr}{\bf\widehat P},$
we characterize the global square bias as follows:\vspace*{-0.5mm}
\bq{btot}
B_{tot}^2\!\!\defeq\!\intli_t\!\intli_f\!\! \left(B(t,f)\!-\!
\sigma_n^2\mbox{tr}{\bf\widehat P}\right)^2\!\! dt\, df
\!=\!\big<{\left|S_{\widetilde P}\right|}^2\!,{\left|EA_x\right|}^2 \big>.
 \vspace*{-1.5mm}
\eq
Just as for the bias we give a global characterization of the
variance. The TF independent term is given by: \vspace*{-1mm}
$$
V_0=\sigma_n^4\|{\bf\widehat P}\|^2 \ . \vspace*{-0.5mm}
$$
We define a total variance as the integral over
the TF dependent variance terms, one has:
\bq{Vtot}
\!\!V_{tot}\!\!\defeq\!\!\intli_t\!\intli_f\!\!\!
 \left(V(t,f)\!-\!V_0\!\right)\!dt\,df
\!=\!\|{\bf\widehat P}\|^2
\!\left(\mbox{tr}{\bf R}_x^2+2\sigma_n^2\mbox{tr}{\bf R}_x\right).
\eq
Equations \refq{btot} and\refq{Vtot} are derived in the appendix.

Observe that {\em any of the global variance constants; i.e.,
the maximum variance $V_{max}$, 
the TF--independent variance term $V_0$, and the total variance
$V_{tot}$ are proportional to the HS norm of the
prototype operator:}  \vspace*{-0.5mm}
\bq{varprop}
V_0,V_{max},V_{tot}\,\propto\,  \|{\bf\widehat P}\|^2 \ .
\eq

\vspace*{2mm}
\section{ESTIMATOR OPTIMIZATION} 

\noindent 
Classical spectrum estimation produces smooth spectra since --- due to
the absence of a model --- smoothing
is the actual tool for variance reduction. The proposed estimators
usually are the result of mean--squared error considerations.
In the present work, we deviate from this point of view in a pragmatic
way: we restrict ourselves to underspread processes whose true spectra
are itself smooth (in the sense of 2D bandlimitation) such that
there exist a whole class of unbiased estimators.  
 While such a modelling
ingredient may be questionable for time--invariant spectrum analysis
we feel that it is necessary for time--varying spectral estimation.
The reason lies in the often overlooked point  that
frequency parametrization  is matched to any stationary process
(the Fourier transform diagonalizes the correlation operator)
while TF parametrization is not matched
to a {\em general} nonstationary process. From the point
of view of operator diagonalization it 
is the class of underspread processes
where TF--parametrization is appropriate \cite{koz94d}.

Unbiased estimation without further assumption on the
signal process $x(t)$ requires a  vanishing ``bias operator'',
i.e., ${\bf P}={\bf\widehat P}$.
In the case of the generalized
Wigner--Ville spectrum,
the prototype operator (cf.~\refq{gwvsp}) 
is not HS since\vspace*{-1mm}
\bq{hsnorm}
\|{\bf P}\|^2=\int_\tau\int_\nu |S_P(\tau,\nu)|^2 d\tau\, d\nu,\vspace*{-1mm}
\eq
so that {\em one can exclude finite--variance unbiased estimation
of the generalized
Wigner--Ville spectrum without a priori knowledge on
the process.}\/ This is well--known  \cite{mar85}.

Based upon the  known support of $EA_x(\tau,\nu)$
one has a large class of nontrivial unbiased estimators (with
nonvanishing ``bias operator'', 
)\vspace*{-1mm}
$$
S^{(\alpha)}_{{\widehat P}_{UB}}(\tau,\nu)= \left\{\begin{array}{r@{,\quad}l}
S^{(\alpha)}_P(\tau,\nu)&\mbox{where}\hspace*{5mm} EA_x(\tau,\nu)\ne0\\[1mm]
                 \mbox{arbitrary}&\mbox{where}\hspace*{5mm} EA_x(\tau,\nu)=0
                 \end{array}\right.. 
$$
We interpret minimum variance in the sense of the combined
consideration of the global variance constants
$V_0,V_{max},V_{tot}$. Due to \refq{varprop}
one has to select the unbiased estimator
with minimum HS norm prototype operator. Using \refq{hsnorm}
this turns out to be
trivial: the minimum--variance unbiased (MVUB)
QTFI estimator is obtained by setting the spreading function of 
the prototype operator zero wherever possible:\vspace*{-2mm}
$$
S^{(\alpha)}_{{\widehat P}_{MVUB}}(\tau,\nu)= \left\{\begin{array}{r@{,\quad}l}
S^{(\alpha)}_P(\tau,\nu)&\mbox{where}\hspace*{5mm} EA_x(\tau,\nu)\ne0\\[1mm]
                 0&\mbox{where}\hspace*{5mm} EA_x(\tau,\nu)=0
                 \end{array}\right.. 
$$
When $\chi_x(\tau,\nu)$ is the smallest indicator function
containing the support of  $EA_x(\tau,\nu)$,
then the MVUB QTFI estimator can be written as:\vspace*{-1mm}
$$
 S^{(\alpha)}_{{\widehat P}_{MVUB}}(\tau,\nu)=
 S^{(\alpha)}_P(\tau,\nu)\chi_x(\tau,\nu) \ .
$$
In particular, for the $\alpha$--parametrized
real--valued generalized Wigner--Ville spectrum one has:\vspace*{-0.5mm}
$$
 S^{(0)}_{{\widehat P}_{MVUB}(\alpha)}(\tau,\nu)=
\cos(2\pi\tau\nu\alpha) \chi_x(\tau,\nu)  \ .
$$
This estimator is optimal among
all QTFI estimators thus in the sense of global variance minimization.
The estimate is locally stable since it minimizes a
bound on the maximum variance ($V_{max}$) and it is unbiased
for arbitrary time and frequency, but
it deviates from the local TF--dependent MVUB estimate. 

\vspace*{2ex}\noindent{\bf Mean--Squared Error.}
The theoretical MVUB estimator serves well as a starting point
for obtaining practical estimators with good mean--squared
error performance. The mean squared error is given by
$E(t,f)=V(t,f)+B^2(t,f)$.  
For any process that satisfies the spreading constraint  \refq{sprecon}
one can formally redefine the estimation target via
the prototype operator of any unbiased estimator:\vspace*{-0.5mm}
$$
P_x(t,f)=\mbox{tr}\left\{{\bf R}_x{\bf P}^{(t,f)}\right\}
=\mbox{tr}\left\{{\bf R}_x{\bf\widehat P}^{(t,f)}_{UB}\right\},
$$
so that one can obtain a useful bound on
the integrated mean--squared error \vspace*{-0.5mm}
\bq{etot}
E_{tot}\!<\!\| {\bf\widetilde P}\|^2\mbox{tr}^2{\bf R}_x +
 {\|{\bf\widehat P}\|}^2\!\!\left(\|{\bf R}_x\|^2+
2\sigma_n^2\mbox{tr}{\bf R}_x\right),
\eq
with ${\bf \widetilde P}={\bf\widehat P}-{\bf\widehat P}_{MVUB}$.
This bound is based on  \refq{btot}, \refq{Vtot} and
$\big<{\left|S_{\widetilde P}\right|}^2\!,{\left|EA_x\right|}^2 
\big>< \|{\bf\widetilde P}\|^2
\mbox{tr}^2{\bf R}_x$. 
 
\section{MATCHED MULTI-WINDOW ESTIMATOR} 

\noindent The eigenfunction decomposition of the  prototype operator ${\bf P}$
shows that  ${\bf P}^{(t,f)}$ is a weighted sum of
rank one projections. Equivalently,   
any QTFI representation
can be written as a weighted sum of spectrograms with
orthonormal windows \cite{she94}.
For practicality, we require our estimator to be based on a  
finite--rank prototype operator with finite--length eigenfunctions.
The MVUB estimator of Sec.~6 does not
satisfy these requirements. Thus, we choose the
finite--rank, time--limited estimator which minimizes
the upper bound on the integrated mean--squared error as given by \refq{etot}.
When we impose the additional requirement that the prototype operator
be projection type with normalized trace, 
${\bf\widehat P}$ has the representation: 
\bq{mm}
{\bf\widehat P}_N=\frac{1}{N}\sum_{k=1}^{N}  \gamma_k\otimes\gamma_k 
\eq
where $ \gamma_k\otimes\gamma_k$ denotes the rank--one projection
on the orthonormal window functions $\gamma_k(t)$ and
$N$ is the rank.
In this case, $ {\|{\bf\widehat P}_N\|}^2 = 1/N$, and the  optimization
of \refq{etot} reduces  to minimizing 
$\|{\bf\widehat P}_{MVUB}\!-\!{\bf\widehat P}_N\|^2$
 subject to orthonormality
constraints on the $\gamma_k$. 
We define the matched multi-window estimator as the quadratic form
based on a prototype operator the minimizes $E_{tot}$ subject
to \refq{mm}. The optimization is performed in a two step procedure:
we optimize the windows subject to a fixed rank 
and then we optimize
the rank. 
For practicality, we impose that the $\gamma_k$ have support on $[-T/2,T/2]$.
To impose this time localization on the optimization of ${\bf\widehat P}_N$,
we define ${\bf T}$ as the projection onto the centered interval 
and require ${\bf\widehat P}_N = {\bf T}{\bf\widehat P}_N{\bf T}$.
Minimizing \refq{etot} yields the optimal windows equation:
\bq{MT}
{\bf T}{\bf\widehat P}_{MVUB}{\bf T}\gamma_{k,opt}=
\lambda_k \gamma_{k,opt}
\eq
The optimum window set is independent of  $N$. 
For the specific case where ${\bf \widehat P}_{MVUB}$ is an ideal bandpass 
(which may be considered as a
 theoretically optimal estimator for stationary processes)
\refq{MT} yields the time--limited and optimally
bandlimited prolate spheriodal wave functions
consistent with \cite{tho82,RST94}.

A more realistic and simpler family of tapers are the discrete sinusoidal
tapers, $\{ v^{(k)}\}$, where
$v_n^{(k)}=\sqrt{\frac{2}{N+1}}\sin\frac{\pi kn}{N+1}$,
and $N$ is the number of points \cite{RS95}. 
The resulting sinusoidal multi-taper spectral
estimate is  $\hat{S}(t,f)=\frac{1}{2K(N+1)}
\sum_{j=1}^K |y(t,f+\frac{j}{2N+2}) -y(t,f-\frac{j}{2N+2})|^2$,
where $y(t,f)$ is the  local Fourier transform 
centered at time $t$ with length $N$.
$S(t,f)$ is the instantanteous spectral density, 
and $K$ is the number of tapers.
The sinusoidal tapers are asymptotically optimal when the bias error is local.

\section{FREE PARAMETER OPTIMIZATION}

For a strongly underspread process $s_x\ll 1$,
$S_P(\tau,\nu)$ is approximately constant in   
 the support of  $EA_x(\tau,\nu)$. Using the optimal window functions
of \refq{MT}, we approximate ${\bf\widehat P}_{MVUB}$ with $1/s_x$ 
such rank one projections. In this case,
$\|{\bf\widehat P}_{MVUB}\!-\!{\bf\widehat P}_N\|^2$ reduces to
$(s_x - 1/N)$. Optimizing \refq{etot} with respect to $N$ 
for moderate noise level yields\vspace*{-0.5mm}
$$
N_{opt}\approx \frac{1}{s_x}, \hspace*{1cm}
\mbox{for}\hspace*{7mm} 
\frac{\sigma_n^2}{\mbox{tr}{\bf R}_x}<\frac{1-s_x}{2}.\vspace*{-2mm}
$$

WERNER: YOUR ESTIMATE of $s_x$ and $T$...

\section{CONCLUSIONS} 

\noindent We have studied time--varying spectral estimation via
quadratic TF--invariant estimators. For circular complex
Gaussian signal and noise processes we have presented explicit
 (local and global) bias and variance results.
 For the specific case of an underspread process
the design of matched multi-window estimators
 has been based on approximating a theoretical MVUB estimator.

The theoretical MVUB estimator as derived in
Section 6 is  a specific case of the
recently proposed optimum kernel design for Wigner--Ville spectrum
estimation \cite{say94}.
We emphasize that  \cite{say94} requires a complete
knowledge  of a second order statistic
what makes this approach purely theoretical
while our proposed estimator uses a more realistic,
incomplete a priori knowledge 
of the process statistics.

For Cohen's class time varying spectra. Using the reproducing kernel
Hilbert space formalism, we derive expressions for the leading order bias
and variance. Underspread processes are band limited in the ambiguity
plane and smooth in the time frequency domain. For underspread processes,
we give unbiased minimum variance estimators.

{\bf APPENDIX: PROOFS}

We now derive \refq{btot} which equates the integral
square bias with the inner product of
the squared GSF of the ``bias operator'' and
the process' expected ambiguity function: \vspace*{-1mm}
{\begin{eqnarray*}
&\D\intli_t\intli_f\mbox{tr}^2\left\{{\bf\widetilde P}^{(t,f)}
{\bf R}_x\right\} dt\,df=\D\intli_t\intli_f
{\left|\big<S_{\widetilde P^{(t,f)}},EA_x\big>\right|}^2dt\,df=&\\
&\D\intli_t\intli_f\intli_{\tau_1}\intli_{\nu_1}\intli_{\tau_2}
\intli_{\nu_2} S_{\widetilde P}(\tau_1,\nu_1)EA_x(\tau_1,\nu_1)
S_{\widetilde P}^*(\tau_2,\nu_2)EA_x^*(\tau_2,\nu_2)&\\
&\cdot \expj{[(\nu_1-\nu_2)t-(\tau_1-\tau_2)f]}
dt\,df\,d\tau_1\,d\nu_1\,d\tau_2\,d\nu_2=&\\&=\D\intli_\tau\intli_\nu
{\left|S_{\widetilde P}(\tau,\nu)\right|}^2
{\left|EA_x(\tau,\nu)\right|}^2 d\tau\,d\nu=
\big<{\left|S_{\widetilde P}\right|}^2,{\left|EA_x\right|}^2\big>&
\ .
\end{eqnarray*}}

Derivation of \refq{Vtot}: 
$$
{\bf P}^{(t,f)}{\bf P}^{(t,f)}=
{\bf S}^{(t,f)}{\bf P}^2{\bf S}^{(t,f)+}=
{\left({\bf P}^2\right)}^{(t,f)}, \vspace*{-1mm}
$$
together with\vspace*{-4.5mm}
$$
\intli_t\intli_f {\bf P}^{(t,f)} dt\, df = 
\mbox{tr}\{{\bf P}\} {\bf I} \vspace*{-2mm}
$$
(which follows directly from the trace formula of the
Weyl correspondence \cite{koz92a}); as well as \vspace*{-2mm}
$$
 \left({\bf P}^{(t,f)}{\bf R}\right)(s,s')=
 \intli_{s''} p(s-t,s''-t)\expjp{f(s-s'')}r(s'',s') ds''\vspace*{-2.5mm}
\ ,$$
whence 
\ba
 \lefteqn{\intli_t\!\intli_f\!\! \mbox{tr}\!\left\{\!\!{\left({\bf P}^{(t,f)}
 {\bf R}\right)}^2\!\right\} dt\, df\!=\!\intli_t\!\intli_f\!
 \intli_s\!\intli_{s'}\!\intli_{s_1}\!\intli_{s_2}\!
 p(s\!-\!t,s_1\!-\!t) r(s_1,s')}\\& &\hspace*{-10mm}
 \cdot p^*(s\!-\!t,s_2\!-\!t) 
  r^*(s_2,s') \expjp{f(s_1-s_2)} dt\,df\, ds\,ds'\, ds_1\,ds_2\\
 &=&\intli_t\!\intli_s\!\intli_{s'}\!\intli_{s_1}
 \left|p(s-t,s_1-t)\right|^2\left|r(s_1,s')\right|^2
  dt\, ds\,ds'\, ds_1\\&=&\|{\bf P}\|^2\|{\bf R}\|^2 \ .
\ea



\end{document}